\documentclass{jfmEdited}
\usepackage{natbib}
\usepackage{graphicx}
\usepackage{epstopdf, epsfig}
\usepackage{xspace}
\usepackage{amsmath}

\newcommand\Ro{\mbox{\textit{Ro}}}
\newcommand\qgp{QG$^{+1}$\xspace} 

\newcommand{\del}{\nabla}
\newcommand\pd{\partial}
\newcommand\qh{\hat{q}}
\newcommand\Fh{\hat{F}}
\newcommand\psih{\hat{\psi}}

\newcommand\bx{\boldsymbol{x}}
\newcommand\bu{\boldsymbol{u}}
\newcommand\bA{\boldsymbol{A}}
\newcommand\bS{\boldsymbol{S}}

\newcommand\buh{\hat{\boldsymbol{u}}}
\newcommand\bAh{\boldsymbol{\hat{A}}}
\newcommand\bSh{\boldsymbol{\hat{S}}}

\shorttitle{Ageostrophic balanced point vortices}
\shortauthor{J. B. Weiss}

\title{\vskip-2cm Point vortex dynamics in three-dimensional ageostrophic balanced flows}

\author{Jeffrey B. Weiss\aff{1} 
  \corresp{\email{jeffrey.weiss@colorado.edu}}
}

\affiliation{\aff{1}Department of Atmospheric and Oceanic Sciences,
  University of Colorado, Boulder, CO 80302 USA}

\begin{document}

\maketitle

\begin{abstract}
Geophysical turbulent flows, characterized by rapid rotation, quantified by small Rossby number, and stable stratification, often self-organize into a collection of coherent vortices, referred to as a vortex gas. The lowest order asymptotic expansion in Rossby number is quasigeostrophy which has purely horizontal velocities and cyclone-anticylone antisymmetry. Ageostrophic effects are important components of many geophysical flows and, as such, these phenomena are not well-modeled by quasigeostrophy. The next order correction in Rossby number, which includes ageostrohpic effects, is so-called balanced dynamics. Balanced dynamics includes ageostrophic vertical velocity and breaks the geostrophic cyclone-anticylone antisymmetry. Point vortex solutions are well known in two dimensional and quasigeostrophic dynamics and are useful for studying the vortex gas regime of geophysical turbulence. Here we find point vortex solutions in fully three dimensional continuously stratified \qgp dynamics, a particular formulation of balanced dynamics. Simulations of \qgp point vortices show several interesting features not captured by quasigeostrophic point vortices including significant vertical transport on long timescales. The ageostrophic component of \qgp point vortex dynamics renders them useful in modeling flows where quasigeostrophy filters out important physical processes. 
\end{abstract}

\section{Introduction}
\label{sec:intro}

Planetary fluids on large scales, often dominated by rapid rotation
and stable stratification, ubiquitously self-organize into coherent vortices. In appropriately scaled vertical coordinates the vortices take the form of roughly spherical patches of potential vorticity \citep{McWilliams1984, McWilliamsEtAl1994, McWilliamsWeiss1994, DritschelEtAl1999, Carton2001, Viudez2016}. A population of physically well-separated coherent vortices, a ``vortex gas'', can be approximated using a hierarchy of idealized models \citep{Kida1981, MelanderEtAl1986, Meacham1992, MeachamEtAl1998,  MiyazakiEtAl2001, DritschelEtAl2004}. The simplest member of the hierarchy is the point vortex model \citep{Helmholtz1867, Kirchhoff1876, Aref2007} in which the vorticity is considered to be concentrated at a collection of isolated points in space. Point vortex models are useful idealizatons for geophysical and astrophysical turbulence and are used to model atmospheres and oceans on the Earth, planets and exoplanets, and astrophysical disks \citep{ArefStremler2001, Carton2001, LucariniEtAl2014, HirtEtAl2018, Abrahamyan2020}. 

The rotation in planetary fluids is characterized by a non-dimensional Rossby number, $\Ro =
U/fL$, the ratio of the planetary rotation timescale $1/f$ to the advective timescale $L/U$, where $U$ is the advective velocity scale, $L$ is a characteristic length scale, $f = 2\Omega \sin(\theta_{lat})$ is the Coriolis parameter, $\Omega$
is the planetary rotation rate, and $\theta_{lat}$ is the latitude. A flow whose advective timescale is much longer than its rotation timescale has $\Ro \ll 1$. 
Large scale midlatitude flows in the Earth's atmosphere typically have $\Ro \sim
0.1$, flows in the Earth's oceans typically have $\Ro \sim 0.01$, while flows on planets and exoplanets often have similarly small Rossby numbers \citep{ShowmanEtAl2010}.
The small Rossby number renders geostrophic balance the dominant  physics: Coriolis forces approximately balance horizontal pressure gradient forces. The leading-order asymptotic theory for rapidly rotating stratified flows is quasigeostrophy (QG), which results from an asymptotic
expansion of the primitive equations for small Rossby number. QG can be derived for flows with many different vertical density structures. Most well-known is QG with a stably stratified background density, either in layers, or with a continuous stratification. Two notable features of stably-stratified QG are that there is no vertical geostrophic velocity and there is an antisymmetry between cyclones and
anticyclones. 

Balance models go beyond QG and include ageostrophic (AG) dynamics as an $O(\Ro)$ correction to QG, retaining the dominant QG balance as the lowest order dynamics. There is a large literature on balance models and many different balance models have been developed, (e.g., \cite{McWilliamsGent1980, Allen1993, Holm1996, MurakiEtAl1999}, and references therein). These models have $O(\Ro^2)$ differences and thus are, formally, equivalently correct through $O(\Ro)$. Different approaches to balance models choose different quantities as a distinguished variable. For example, \citep{Allen1993} chooses pressure. Here we investigate the balance model of \cite{MurakiEtAl1999} (MSR1999), \qgp, which chooses potential vorticity $q$ as its distinguished variable. This makes it a natural choice for formulating balanced point vortices. Balance models differ qualitatively from QG in that the AG dynamics includes a
nonzero $O(\Ro)$ vertical velocity and breaks the cyclone-anticyclone
antisymmetry. Despite being small, both of these effects are important for many aspects of
planetary fluid dynamics. There have been a number of studies of finite-sized coherent vortices in balance models as well as in primitive equation models with varying Rossby number (e.g. \cite{TsangDritschel2015, McKiverDritschel2016, MahdiniaEtAl2017, ReinaudDritschel2018, McKiver2020,SokolovskiyEtAl2020}, and references therein). 

Point vortex models have a long history in two-dimensional (2d) fluid dynamics (see \cite{Aref2007} for a review). They are also well developed in layered and continuously stratified QG dynamics \citep{Gryanik1983, Gryanik1991,GryanikEtAl2000, GryanikEtAl2006,ReznikKizner2007}.  
Due to the constraints of QG dynamics, QG point vortex models are limited to modeling physical situations where the vertical velocity and cyclone-anticyclone antisymmetry are unimportant. 

Here we find point vortex solutions to the nonlinear balance
model of MSR1999, \qgp. 
The solutions have many similarities to 2d and QG point vortices, but go beyond them by inheriting from balance models $O(\Ro)$ vertical velocities and
broken cyclone-anticyclone antisymmetry. In section~\ref{sec:qgp1}, we
review the aspects of \qgp theory relevant for the development of \qgp point vortex dynamics. In section~\ref{sec:vortex-gas} we explore vortex-gas solutions of \qgp. Section~\ref{sec:ff} finds the equations of motion for \qgp point vortices advected by the far-field of the other vortices, section~\ref{sec:self} shows that the self-advection is zero in the point vortex limit, and section~\ref{ref-summary} summarizes the \qgp point vortex equations of motion. Section~\ref{sec:properties} describes some of the properties of \qgp point vortex dynamics obtained by analyzing the equations of motion. Section~\ref{sec:trajectories} briefly describes a few numerical simulations of the \qgp point vortex equations, highlighting some of the new features of \qgp point vortices.

\section{\qgp Theory}
\label{sec:qgp1}

Following MSR1999, we begin with an inviscid, adiabatic,
three-dimensional Boussinesq hydrostatic fluid on an $f$-plane (i.e. constant Coriolis parameter
$f$) described by the primitive equations,
\begin{align}
\del\cdot\bu &= 0, 
\label{eq:peinc}\\
\frac{D \bu_h}{Dt} + f \hat{z} \times \bu_h &= -\del_h \phi^T,
\label{eq:pehmom}\\
\frac{D\theta^T}{Dt} &= 0,
\label{eq:peenergy}\\
\partial_z\phi^T &= g \theta^T/\theta_0,
\label{eq:pehydro}\\
\frac{D}{D t} &= \partial_t + \bu\cdot\del,
\label{eq:peadv}
\end{align}
where $\bu = u\hat{x}+ v\hat{y} + w\hat{z}$ is the three-dimensional (3d) fluid velocity,  $\bu_h = u\hat{x}+ v\hat{y}$ is the horizontal velocity, $\del$ and $\del_h$ are the 3d and horizontal gradient operators, respectively, $\theta^T$ is the total potential temperature, and $\theta_0$ is a reference potential temperature. We follow MSR1999 and refer to $\phi^T$ and $z$ as the total pressure and vertical height, although strictly speaking they are more properly a modified pressure and the geopotential height \citep{HoskinsBretherton1972}. Eqs.~\ref{eq:peinc}-\ref{eq:peadv} represent, in order, incompressibility, conservation of horizontal momentum, conservation of energy, hydrostatic balance, and the definition of the three-dimensional material derivative. 

We restrict ourselves to
flows with a constant Brunt-V\"ais\"al\"a frequency $N$.
The pressure and potential temperature are separated into horizontally uniform reference fields $\phi^{ref}$ and $\theta^{ref}$, and departures from the reference fields
\begin{align}
    \phi^{ref}&= \frac{1}{2} N^2 z^2,\\
    g \theta^{ref}/\theta_0 &= N^2 z,\\
    \phi^T &= \phi + \phi^{ref}, \\
        \theta^T &= \theta + \theta^{ref}.
\end{align}

The dynamics is nondimensionalized as in MSR1999 in terms of a vertical distance $H$, a horizontal distance $L$, and a velocity $U$. Two non-dimensional parameters arise: the Burger number which is taken to be unity, $B = (NH/fL)^2=1$, and the Rossby number which is assumed to be small, $Ro= U/fL \ll 1$. These scalings give the non-dimensional primitive equations. 

Three scalar fields have explicit advective dynamics: the horizontal velocity vector and the potential temperature. The remaining variables, the pressure and vertical velocity, are determined diagnostically from incompressibility and hydrostatic balance, respectively. MSR1999 write the dynamics in terms of a 3d vector potential $\bA$, with $\bu = \nabla\times\bA$ and $\theta = -\nabla\cdot\bA$. The curl relation for the velocity guarantees that the fluid is incompressible in 3d. 

The relevant potential vorticity is the Ertel potential vorticity, $Q = (\hat{z} + Ro \del\times\bu_h)\cdot\del\theta^T$. The dynamically active disturbance potential vorticity $q$ is the departure from a uniform potential voriticy which, in this scaling, is unity, $Q = 1 + Ro \, q$. Following MSR1999, we refer to this disturbance Ertel potential vorticity $q$ as just the potential vorticity (PV). 

Technically, QG appears at $O(\Ro)$ in the full asymptotic expansion. As seen, for example, in the relation between $Q$ and $q$ above, the $O(1)$ term is uninteresting and is typically removed, following which the QG quantities are scaled by $\Ro$ and are rendered $O(\Ro^0)$. At next order, ageostrophic (AG) effects enter. QG and AG fields will be denoted, respectively, by the subscripts ``0'' and ``1''. The asymptotic expansions then take the form, using the vector potential as an example, of
\begin{equation}
    \bA = \bA_0 + Ro \,\bA_1 + O(Ro^2).
\end{equation}


The familiar QG relations between horizontal velocity, PV, and streamfunction,
\begin{equation}
    \bu = \begin{pmatrix}
    -\partial_y\psi_0\\ \hphantom{-}\partial_x \psi_0 \\ 0
    \end{pmatrix},
    \qquad q_0=\nabla^2\psi_0,
\end{equation}
are rewritten in terms of the vector potential which is the solution to a Poisson equation $\nabla^2 \bA_0 = \bS_0$ where $\bS_0$ is the QG source,
\begin{equation}
    \bA_0 = \begin{pmatrix}
    0\\0\\-\psi_0
\end{pmatrix}
\qquad
\bS_0 = \begin{pmatrix}
0\\0\\-q_0
\end{pmatrix}.
\end{equation}
This structure carries over to higher order $\bu_n=\nabla\times\bA_n$, $\nabla^2 \bA_n = \bS_n$, where, unlike the QG terms, the higher order $\bA_n$ $\bS_n$ are generally nonzero in all three vector components.

The \qgp model is a specific instance of an iterated balance model \citep{Allen1993}. The iteration procedure is based on treating one distinguished  physical variable as exact and iteratively computing corrections to other variables at each instant of time. 
For example, \cite{Allen1993} chose pressure as the distinguished variable. \qgp chooses PV which makes it natural to seek balanced point vortex solutions in \qgp. Other balance models may also have point vortex solutions but we defer that question to future work.

From a vorticity perspective, QG flow is completely described by the QG PV, $q_0(\bx,t)$. The QG streamfunction is obtained by solving a Poisson equation with $q_0$ as the source, which then determines the velocity and potential temperature. 
Conservation of QG potential vorticity under QG flow is given by
\begin{equation}
\frac{D_0 q_0}{D t} = \partial_t q_0 + \left(\bu_0\cdot\nabla\right) q_0 = 0,
\end{equation}
where $D_0/Dt$ indicates the material derivative of fluid parcels
advected with the QG velocity.

\qgp dynamics is governed by potential vorticity conservation, where the full potential vorticity $q$ is advected by the approximate velocity
\begin{equation}
\label{eq:pvcons}
    \frac{D q}{Dt} = \frac{\pd q}{\pd t} + \left ( (\bu_0 + \Ro\  \bu_1 + O(\Ro^2))\cdot\del\right) q = 0
\end{equation}
The potential vorticity can be partitioned into QG and AG contributions, $q = q_0 + \Ro\ q_1 + O(\Ro^2)$, however, as discussed in
MSR1999, this partition has some subtleties. For our
purposes, we will show that for point vortices, $q_1 = 0$ is consistent with \qgp and proceed assuming $q = q_0$.

MSR1999 showed that the AG source $\bS_1$ is an operator acting on the QG streamfunction $\psi_0$ plus the vector  $(0,0,-q_1)$. With our assumption that $q_1 = 0$, the AG source becomes
\begin{align}
\bS_1
&= 
\begin{pmatrix}
-2 J\left(\partial_z\psi_0,
\partial_y\psi_0\right)\\
\hphantom{-}2 J\left(\partial_z\psi_0,
\partial_x\psi_0\right) \\
\left(\nabla^2 \psi_0\right)\partial_{zz} \psi_0 
- \left| \nabla \left(\partial_z \psi_0\right)\right|^2 
\end{pmatrix},
\label{eq:potS1}
\end{align}
where $J$ is the usual horizontal Jacobian operator, $J(f,g) =
(\partial_x f) (\partial_y g) - (\partial_y f)(\partial_x g)$.  Depending on the context, it is convenient to consider $\bS_1$ as either a function of position $\bS_1(\bx)$ obtained from considering $\psi_0(\bx)$ as an explicit function of position, or as a differential operator acting on a function $\psi_0$. Since $\psi_0$ appears quadratically in $\bS_1$, we can, in the operator view, consider $\bS_1$ to an operator acting on two potentially independent scalar functions $f$ and $g$,
\begin{equation}
\bS_1
= 
\begin{pmatrix}
-2 J\left(\partial_z f,
\partial_y g\right)\\
\hphantom{-}2 J\left(\partial_z f,
\partial_x g\right) \\
\left(\nabla^2 f\right)\partial_{zz} g 
-  \left(\nabla \partial_z f\right)\cdot \left(\nabla \partial_z g\right) 
\end{pmatrix}.
\end{equation}
In this perspective, $\bS_1$ is a bilinear non-symmetric vector differential operator with the properties
\begin{align}
    &\bS_1(c_1 f_1+c_2 f_2,g) = c_1\bS_1(f_1,g)+c_2 \bS_1(f_2,g),\\
    &\bS_1(f, c_1 g_1 + c_2 g_2) = c_1\bS_1(f,g_1)+c_2 \bS_1(f,g_2),\\
    &\bS_1(f,g) \neq \bS_1(g,f),
\end{align}
where $f$'s and $g$'s are scalar functions of $\bx$ and $c$'s are constants. The bilinearity of $\bS_1$ plays a large role in the subsequent development of \qgp vortex gas dynamics where $\psi_0$ is decomposed into a sum of contributions from individual vortices. Note 
that since the AG vector potential, through the AG source, depends quadratically on the QG
streamfunction, it manifestly breaks the cyclone-anticylone (anti)symmetry of QG,
\begin{align}
    &\bA_0 \xrightarrow[q_0 \to 
-q_0]{} -\bA_0, \\
 &\bA_1 \xrightarrow[q_0 \to 
-q_0]{} +\bA_1,
\end{align}
and so $\bA = \bA_0 + \Ro \bA_1$ is non-symmetric under sign changes of the PV.

MSR1999 consider a fluid with doubly-periodic horizontal boundaries and solid vertical boundaries. It is usual to consider point vortex dynamics in an infinite domain with boundary conditions requiring physical fields to decay to zero as one goes to infinity. Then, point vortex dynamics in closed and periodic domains are obtained through the method of images. Here we follow the point vortex viewpoint and consider \qgp in a three-dimensional infinite domain and require that the velocity and potential temperature go to zero at infinity.

The \qgp equations are completed by solvability constraints, which, in the infinite domain consider here, are
\begin{align}
\iiint q_0 \ d^3\bx &= \lim_{z\to\infty}\iint
\left.\theta_0\right|^{+z}_{-z} dx dy, \label{eq:solve1}\\
\iiint q_1 \ d^3\bx &= \lim_{z\to\infty}\iint \left[\left.
\theta_1 + \pd_z \psi_0\left(\pd_{xx}\psi_0 +
\pd_{yy}\psi_0\right)\right]\right|^{+z}_{-z} dx dy, \label{eq:solve2}
\end{align}
where equation~\ref{eq:solve1} guarantees the invertibility of the QG
potential vorticity-streamfunction relation, and Eq~\ref{eq:solve2}
guarantees the invertibility of the $z$-component of equation~\ref{eq:potS1}. 


\section{The vortex gas approximation} 
\label{sec:vortex-gas}

We begin by assuming the fluid takes the form of a so-called ``vortex gas'', where the PV consists of a collection of $N$ physically separated coherent vortices. Each vortex has a location, $\bx_i(t)$, and a circulation $\Gamma_i$. We assume each coherent vortex has the same finite-sized axisymmetric shape in PV, $\qh$. The only time dependence in the flow is in the location of the vortices. The number of vortices, their shapes, and their circulations are assumed to be constant. The PV field is then 
\begin{equation}
\label{eq:cvsum}
  q(\bx) = \sum_{i=1}^N \Gamma_i\qh(\bx - \bx_i(t)),
\end{equation}
where $\qh$ is normalized to have unit circulation, $\int \qh \, d^3\bx = 1$. Potential vorticity conservation, equation~(\ref{eq:pvcons}), then requires the vortex locations move with the local velocity,
\begin{equation}
\label{eq:pvadvection}
    \frac{d \bx_i(t)}{dt} = \bu(\bx_i) = \bu_0(\bx_i) + \Ro\  \bu_1(\bx_i) + O(\Ro^2),
\end{equation}
where we have used
\begin{equation}
\label{eq:dqdt}
    \frac{\pd \qh(\bx - \bx_i(t))}{\pd t} = - \left. \nabla\qh(\bx)\right|_{\bx = \bx - \bx_i}\cdot \frac{d \bx_i(t)}{dt}
\end{equation}
Thus, while conservation of the full $q$ in \qgp gives AG corrections to the motion of the vortices, the vortex itself is advected coherently.

The logic of \qgp, reflecting the iterated nature of balance models generally, is that one begins with the QG PV $q_0$, uses the QG relations to obtain the QG streamfunction $\psi_0$, and then uses $\psi_0$ to obtain the AG fields. Following this logic for a  vortex gas, the fields decompose into sums over shape fields, denoted by hats, which are identical for all vortices. Denoting the asymptotic order by the subscript $n=0$ for QG and $n=1$ for AG, and letting $F$ represent any of the fields $\bS$, $\bA$, $\bu$, and $\theta$, vortex-gas fields take the form
\begin{align}
    F_0(\bx, t) &= \sum_{i=1}^N \Gamma_i \Fh_0(\bx - \bx_i),\\
    F_1(\bx, t) &= \sum_{i, j =1}^N \Gamma_i \Gamma_j \Fh_1(\bx - \bx_i, \bx - \bx_j).
    \label{{eq:singlepairF}}
\end{align}
The QG shape fields $\Fh_0$ are single-vortex fields and depend on the location of a single vortex, while the AG  shape fields $\Fh_1$ are pair-vortex fields and depend on the locations of a pair of vortices. The pair-vortex form of the AG fields is a consequence of both the vortex-gas approximation and the bilinear nature of the AG source operator.
This vortex-pair interaction is fundamentally different from the vortex pair dynamics of 2d and QG point vortices. In 2d and QG dynamics, all fields are single-vortex fields and the interaction is solely through the sum of the advection induced by each vortex separately. In \qgp, vortex pairs interact in creating the AG source, $\bS_1$, for the AG vector potential, $\bA_1$, and the advecting velocity is the sum of velocities induced by vortex pairs. At next order in $\Ro$, three-vortex and four-vortex contributions would enter. 

The AG fields can be rewritten in terms of  single-vortex nonlinear contributions $\Fh_1^{s}$, and symmetrized pair-vortex nonlinear contributions $\Fh_1^p$
\begin{align}
    \Fh_1^{s}(\bx - \bx_i) &= \Fh_1(\bx - \bx_i, \bx - \bx_i),\\
     \Fh_1^{p}(\bx - \bx_i, \bx - \bx_j) &=  \Fh_1(\bx - \bx_i, \bx - \bx_j) + 
     \Fh_1(\bx - \bx_j, \bx - \bx_i)
    \label{{eq:pairF}}
\end{align}
resulting in
\begin{equation}
    F_1(\bx,t) = \sum_{i=1}^N \Gamma_i^2 \Fh_1^{s}(\bx - \bx_i) + \sum_{\substack{i=1\\j=i+1}}^N \Gamma_i \Gamma_j\Fh_1^{p}(\bx - \bx_i, \bx - \bx_j).
\label{eq:Fsum}
\end{equation}
In all sums we use the convention that terms in a sum that contradict the bounds, such as the term in the second sum above with $i=N$ and $j=i+1 = N+1$, are zero.

Following the logic of a \qgp vortex gas, all the shape fields are determined by the potential vorticity shape field $\qh$ through solving Poisson equations and applying differential operators:
\begin{align}
& \bSh_{0}(\bx) = \begin{pmatrix}
0\\
0\\
-\del^2\psih_0(\bx)
\end{pmatrix}=\begin{pmatrix}
0\\
0\\
-\qh(\bx)
\end{pmatrix}\\
& \bSh_1^s(\bx - \bx_i) =  \bS_1(\psih_0(\bx - \bx_i),\psih_0(\bx - \bx_i))\\
& \bSh_1^p(\bx - \bx_i, \bx - \bx_j) =  \bS_1(\psih_0(\bx - \bx_i),\psih_0(\bx - \bx_j)) + i \longleftrightarrow j\\
& \del^2 \bAh_n^{\{s,p\}} = \hat{\bS}_n^{\{s,p\}},\\
& \buh_n^{\{s,p\}} =  \del\times \bAh_n^{\{s,p\}},
\end{align}
where the operator $\bS_1$ is given by equation~(\ref{eq:potS1}).

Vortex-gas advection, equation~(\ref{eq:pvadvection}), becomes 
\begin{multline}
\label{eq:asympadv}
\frac{d \bx_i(t)}{dt} = \sum_{j=1}^N \Gamma_j \buh_0(\bx_i- \bx_j) \\
+ \Ro \left(\sum_{j=1}^N \Gamma_j^2  \buh_1^s(\bx_i-\bx_j) + \sum_{\substack{j=1\\k=j+1}}^N \Gamma_j \Gamma_k \buh_1^p(\bx_i-\bx_j, \bx_i - \bx_k)\right).
\end{multline}
The advecting velocities fall into two categories: far-field advection, where the single vortex $j$ and the pair of vortices $j, k$ are different than the advected vortex $i$, and self-advection, where either $i = j$ or $i = k$. It is well known that 2d and QG vortices do not self-advect on the $f$-plane considered here, $\buh_0(0) = 0$ (vortices on the $\beta$-plane do, however, self-advect). Below we will show that in addition there is no AG self-advection, $\buh_1^s(0) = \buh_1^p(0,\bx) = \buh_1^p(\bx,0) = 0$.

\section{\qgp Point Vortex Dynamics}
\label{sec:pvdynamics}

\subsection{Far-field advection}
\label{sec:ff}

The far-field vector potentials can be determined by taking the vortex size to zero before performing the calculations,  $\qh = \delta(\bx)$. We assume for now that $q_1 = 0$, i.e., the QG point vortex carries all the potential vorticity. We will show below that this choice is consistent with the solvability constraint equation~(\ref{eq:solve2}). To solve for  single-vortex fields we can assume the vortex is at the origin. 

As the QG vortex shape is a $\delta$-function, the streamfunction shape, solving $\nabla^2\psih_0 = \delta(\bx)$, is the Green's function, $\psih_0 = G(\bx) = -1/4 \pi |\bx|$, and the QG vector potential shape is $\bAh_0 = (0,0, -\psih_0)$.  
The single-vortex AG Poisson source $\bSh_1^s$ from a vortex at the origin can be directly calculated, leading to
\begin{equation}
\bSh_1^s = \frac{1}{8 \pi ^2 |\bx|^8}
\begin{pmatrix}
3  x z \\
3  y z \\
-\left(x^2+y^2+4
  z^2\right)/2 \\
\end{pmatrix}\label{eq:Sexplicit}
\end{equation}

We solve the Poisson problem by first seeking a solution that matches the denominator of the source. We note that 
$\nabla^2 |\bx|^{-m} \sim |\bx|^{-m-2}$. Since the
denominator of $\bSh_1^s$ is proportional to $|\bx|^{8}$, we
expect the denominator of $\bAh_1^s$ to be proportional to
$|\bx|^{6}$. Dimensionally,  
$\bSh_1^s \sim 1/L^6$, so $\bAh_1^s \sim 1/L^4$. This means the numerator of
$\bAh_1^s$ must be quadratic in $(x, y, z)$. We thus set the numerator of each
component of $\bAh_1^s$ to a general quadratic $a x^2 + b y^2 + c z^2 + d
x y + e x z + f y z$, and take the Laplacian of this general $\bAh_s^2$. Requiring that $\bAh_1^s$
solves the Poisson equation determines the values of the constants in the quadratic. The
result is 
\begin{equation}
\bAh_1^s(\bx) = \frac{z}{16 \pi^2 |\bx|^6}
\begin{pmatrix}
x\\y\\-z/2\\
\end{pmatrix}\label{eq:A1ssoln}
\end{equation}

The next step is to solve the Poisson equation for the far-field pair-vortex potential. We notice that the components of the far-field single-vortex potential (\ref{eq:A1ssoln}) can each be written as the product of two functions
\begin{equation}
    \bAh_1^s = \begin{pmatrix}
h_x(\bx) h_z(\bx) \\ h_y(\bx) h_z(\bx) \\ - h_z(\bx)^2/2
\end{pmatrix},
\end{equation}
where $\boldsymbol{h} = \bx/4 \pi |\bx|^3 = \nabla G(\bx)$  and its subscript represents the corresponding vector component. We guess that $\bAh_1^p$ takes a similar symmetrized form
\begin{equation}
    \bAh_1^p(\bx-\bx_i, \bx-\bx_j) = 
    \begin{pmatrix}
h_x(\bx-\bx_i) h_z(\bx-\bx_j) + h_x(\bx-\bx_j) h_z(\bx-\bx_i) \\ h_y(\bx-\bx_i) h_z(\bx-\bx_j) + h_y(\bx-\bx_j) h_z(\bx-\bx_i) \\ 
-h_z(\bx-\bx_i) h_z(\bx-\bx_j)
\end{pmatrix}.\label{eq:A1psoln}
\end{equation}
Direct calculation of $\del^2 \bAh_1^p$ verifies that this is indeed the solution. These solutions manifestly satisfy the boundary conditions, going to zero at infinity.

The right-hand-side of the integral constraints involve integrals evaluated on the vertical boundaries, which we take here to be at $z = \pm \infty$. We assume all vortices are at finite heights, and then the integral constraints only depend on the far-fields. The QG integral constraint equation~(\ref{eq:solve1}) can be integrated analytically and verified. We have not found an analytic form for the integral on the RHS of equation~(\ref{eq:solve2}). However, the scaling behavior of the fields lets us deduce that the integral is zero. Since $\bAh_1$ decays as $|\bx|^{-4}$, $\hat\theta_1$ decays as $|\bx|^{-5}$. $\hat\psi_0$ decays as $|\bx|^{-1}$, so the second term inside the integral also decays as $|\bx|^{-5}$. The RHS thus goes to zero as $z\to\pm\infty$, indicating that the integral of $q_1$ is zero, consistent with $q_1 = 0$.

\subsection{Self-advection}
\label{sec:self}

Point vortex self-advection is the limit of the finite-sized coherent vortex self-advection as the vortex size goes to zero. This requires some consideration of a finite vortex. Here we consider all vortices to have the same time-independent, spherically symmetric shape with size $r_0$. 

In 2d and QG point vortex dynamics, the self-advection velocity of vortex $i$ is the velocity at $\bx$ induced by a single finite-sized coherent vortex in the ordered limit, $\lim_{r_0\to0}\lim_{\bx\to\bx_i}$.
In \qgp point vortex dynamics, there are three self-advection velocities: $\buh_0(0)$ the QG velocity at a vortex center induced by its own vorticity distribution;
$\buh_1^s(0)$, the AG velocity at a vortex center induced by the nonlinear interaction between the vortex and itself; and $\buh_1^p(0,\Delta\bx)$, the AG velocity at a vortex center induced by the nonlinear interaction between the advected vortex and a different vortex separated by $\Delta\bx$. 

The consequences of the assumed vortex shape symmetry are leveraged using the 3d infinite-domain Green's function, written as
\begin{equation}
    G(\bx,\bx') = \frac{-1}{4\pi |\bx - \bx'|},
\end{equation}
and the Green's function solution of the Poisson equation,
\begin{align}
\del^2 \bA_n(\bx) = \bS_n(\bx) \,\Longleftrightarrow \,\bA_n(\bx) = \int G(\bx, \bx') \bS_n(\bx') dV'.
\end{align}
The velocity is then
\begin{equation}
    \bu_n(\bx) = \del \times \bA_n(\bx) = \int \del_{\bx} G(\bx,\bx') \times \bS_n(\bx') d^3\bx',
\end{equation}
where the subscript on the gradient operator indicates derivatives are with respect to $\bx$.

We first recall the QG point vortex self-advection, $\buh_0(0)$, which can be calculated by considering a single unit-circulation vortex at the origin. Then
\begin{equation}
\label{eq:u0self}
\buh_0(0) = \lim_{r_0\to0}\int [\del_{\bx} G(\bx,\bx')]_{\bx = 0} \times \bSh_{0}(\bx') d^3\bx'.
\end{equation}
The symmetries in the vortex shape and the Green's function render the argument of the integral in equation~(\ref{eq:u0self}) odd in one direction and thus $\buh_0(0) = 0$, there is no QG point vortex self-advection.

The AG single-vortex self-advection $\buh_1^s(0)$ is also zero by symmetry but the argument is a bit more complicated. The relevant source is $\bSh_1^s(\bx')$. The derivatives in equation~(\ref{eq:potS1}), allow one to determine the symmetry of the components of $\bSh_1^s(\bx')$ in the different directions. One can show that each component has at least one direction where the integrand is odd and integrates to zero. Thus, $\buh_1^s(0) = 0$ and there is no AG single-vortex self-advection.

Physically, we know that single coherent vortices on the $f$-plane do not self-advect and so it is satisfying that symmetry-based arguments show no single-vortex self-advection. The pair self-advection is different. Since a pair of QG vortices do advect each other, there is no physical reason why the AG pair self-advection must be zero. A finite AG pair self-advection would just provide a small correction to the QG mutual advection and is not a priori ruled out on physical grounds.

Because the pair self-advection relies on a second vortex at location $\bx_2 \ne 0$, one cannot use symmetry arguments to determine its value. Instead we explicitly consider a specific finite vortex shape. One common finite vortex shape is a Gaussian monopole
\begin{align}
    \qh^{r_0} &= \frac{e^{-r^2/2 r_0^2}}{(2\pi r_0^2)^{3/2}}, \\
    \psih_0^{r_0} & = G(r) \hbox{Erf}(r/(\sqrt{2} r_0)),
\end{align}
where $r = |\bx|$, $\hbox{Erf}(s) = \frac{2}{\sqrt{\pi}}\int_0^s e^{-t^2}\,dt$ is the error function, and the superscript $r_0$ denotes a finite vortex.

To calculate the AG pair self-advection of a vortex, we can place that vortex at the origin, $\bx_1 = 0$. We are interested in the velocity near the vortex center, the origin. In the vortex-gas approximation, the second vortex is far away $|\bx_2| \gg r_0$, and we can use its far-field streamfunction $\psih_0 = G(\bx - \bx_2)$ in calculating $\bSh_1^p$:
\begin{equation}
    \bSh_1^p = \bS_1(\psih_0^{r_0}(\bx), \psih_0(\bx - \bx_2)) + \bS_1( \psih_0(\bx - \bx_2),\psih_0^{r_0}(\bx)) 
\end{equation}
Expanding $\bSh_1^p$ in a Taylor series near the origin, $r \ll r_0$, gives $\bSh_1^p\sim  O(r^0)$. This then implies that $\bAh_1^p \sim O(r^2)$ and  $\buh_1^p \sim O(r)$ near the origin. Thus $\buh_1^p \to 0$ as $r \to 0$ for finite Gaussian vortices and remains zero in the point vortex limit $r_0\to 0$. Thus, at least for the point vortex limit of Gaussian vortices, the AG pair self-advection is zero.

\subsection{Summary equations}
\label{ref-summary}
As shown above, \qgp point vortices, like 2d and QG point vortices, have no self-advection. Thus, their dynamics is completely determined by their far-field advection. Putting together the pieces described above gives the equations of motion for a set of $N$ \qgp point vortices:
\begin{align}
    \frac{d \bx_i}{dt} &= \bu(\bx_i),\label{eq:dxdt}\\
    \bu(\bx_i) &= \sum_{j=1}^N {}^{'} \left( \Gamma_j\buh_0(\bx_i - \bx_j) + \Ro\ \Gamma_j^2 \buh_1^s(\bx_i - \bx_j)\right)  \nonumber\\ 
    &\qquad + \Ro \sum_{\substack{j=1\\k=j+1}}^N{}^{'} \Gamma_j \Gamma_k \buh_1^p(\bx_i - \bx_j, \bx_i - \bx_k) \label{eq:upsum}
\end{align}
where $\sum{}^{'}$ denotes skipping terms with $j = i$ and $k = i$. The single-vortex and pair-vortex velocities are
\begin{align}
    \buh_0(\bx) &= \frac{1}{4 \pi |\bx|^3}
    \begin{pmatrix}
        -y\\ x\\ \,0
    \end{pmatrix},
    \label{eq:u0hat}\\
    \buh_1^s(\bx) &= \frac{x^2+y^2-8z^2}{16 \pi^2 |\bx|^8}
    \begin{pmatrix}
        -y\\ x\\ \,0
    \end{pmatrix},\label{eq:u1hats}\\
    \buh_1^p(\bx_1,\bx_2) &= \frac{\tilde{\bu}}{16 \pi^2 |\bx_1|^5 |\bx_2|^5},\label{eq:u1hatp}
\end{align}
where
\begin{equation}
    \tilde{\bu} = 
    \begin{pmatrix}
    3 |\bx_1|^2 (y_1 z_2^2 + 2 y_2 z_1 z_2) 
        + 3|\bx_2|^2(y_2 z_1^2 + 2 y_1 z_1 z_2)  - |\bx_1|^2 |\bx_2|^2 (y_1+y_2) \\
         |\bx_1|^2 |\bx_2|^2 (x_1+x_2)- 3 |\bx_1|^2 (x_1 z_2^2 + 2 x_2 z_1 z_2) 
        - 3|\bx_2|^2(x_2 z_1^2 + 2 x_1 z_1 z_2)\\
         3 (x_2 y_1 - x_1 y_2)( |\bx_2|^2 z_1 - |\bx_1|^2 z_2).
    \end{pmatrix} \label{eq:upnum}
\end{equation}
We note that, as is the case for 2d and QG point vortices, it is straightforward to study passive scalar transport by considering passive tracers as vortices with zero circulation.

Asymptotic expansions break down if the higher order terms become larger than lower order terms. Here, asymptotic consistency requires  $|\buh_1^s| /|\buh_0| \sim O(1)$ and $|\buh_1^p| /|\buh_0| \sim O(1)$, which requires $\Gamma/4\pi|\bx|^3 \sim O(1)$. For single-vortex advection, $\bx$ is the distance from the advected particle, either a vortex or passive, to the advecting vortex and $\Gamma$ is the circulation of the advecting vortex. For pair-vortex advection, $\bx$ is the distance to the nearest member of the advecting pair and $\Gamma$ is its circulation. For a collection of point vortices with similar circulations, $\Gamma_i \sim O(\Gamma)$, there is thus a minimum distance between \qgp vortices and between passives and vortices, an``asymptotic horizon'', if you will, $r_a \sim (\Gamma/4\pi)^{1/3}$. The asymptotic horizon is a boundary in phase space and inside the boundary the asymptotic expansion becomes misordered. Systems where all vortices remain further apart than $r_a$ along their entire trajectories are outside the horizon and are asymptotically valid \qgp solutions. Similarly, a passive particle which remains further than $r_a$ from all vortices is an asymptotically valid passive trajectory. There is no restriction on how close passives can approach each other as they do not induce any advective velocity. For systems with a wide range of $\Gamma$'s, the asymptotic horizon becomes more complex. 

\section{Properties of \qgp point vortex dynamics}
\label{sec:properties}

\subsection{vortex pairs}
The qualitative behavior of a system composed of only two vortices, $N=2$ can be understood from the dynamical equation, Eqs.~(\ref{eq:dxdt})-(\ref{eq:upnum}). We begin by noting that for two vortices, it is not possible to have $i$, $j$, $k$ which satisfy $i\ne j$, $i\ne k$, and $k>j$. As a result, the second sum in equation~(\ref{eq:upsum}) has no terms and the pair-vortex velocity is zero. 

Both 2d and QG vortex pairs co-rotate in horizontal planes. Since $\buh_0$ and $\buh_1^s$ have zero vertical velocity and $\buh_1^p=0$, \qgp vortex pairs also move in horizontal planes with no vertical velocity. $\buh_1^s$ shares with $\buh_0$ the property that it is proportional to $(-y, x, 0)$ and is thus directed perpendicular to the line connecting the vortices. The AG velocities induced at vortices $1$ and $2$ are, as in 2d and QG, in opposite directions. Thus, \qgp vortex pairs also co-rotate. 

The AG correction can either speed up or slow down the QG rotation. The magnitude of the AG velocity scales as $\Gamma^2$ and so is positive for both same-sign and opposite sign vortex pairs. Furthermore, the numerator of $\buh_1^s$ changes sign on the surfaces $z = \pm \sqrt{(x^2+y^2)/8}$. As \qgp vortex pairs rotate with constant vertical and horizontal separations, the sign of the numerator is constant. Thus, whether \qgp vortex pairs rotate faster or slower than QG vortex pairs depends on both the signs of the circulations and the specific vector vortex separation. The difference in rotation speeds with the sign of the vortex circulation is one example of \qgp breaking the cyclone-anticylone anti-symmetry.

The relative magnitudes of the AG and QG vortex speeds, $|\bu_1^s|/|\bu_0|$ scales as $\Ro\Gamma/{|\bx_1-\bx_2|^3}$. Thus, as the vortex separation grows, the AG velocity decays much faster than the QG velocity.

\begin{figure}
  \centering
  \includegraphics[width=4cm]{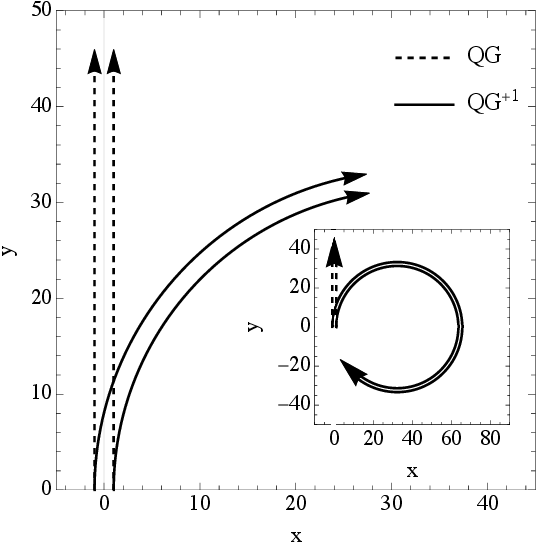}
  \caption{QG (dashed) and \qgp (solid) heton trajectories viewed from above for initial vortex positions $\pm(1, 0, 1)$ with $\Gamma = \mp 4\pi$ and $\Ro = 0.2$. The height and separations of the vortices comprising QG and a \qgp heton are constant. The main figure shows the horizontal projection of the trajectories for $0 \le t \le 500$ and the inset shows the projection for the \qgp heton for $0 \le t \le 2000$}
  \label{fig:Heton}
\end{figure}

Opposite sign 2d and QG vortices with $\Gamma_1 = - \Gamma_2$, called hetons in the oceanographic context, propagate along straight lines. Due to the dependence of the AG velocity on $\Gamma^2$, the two vortices no longer have the same horizontal velocity and they travel in circles, similar to opposite-sign unequal magnitude vortices in 2d and QG, with a radius of curvature that scales as $1/\Ro$ (figure \ref{fig:Heton}). The heton curvature breaks the cyclone-anticyclone anti-symmtery. If the symmetry held, switching the signs of the circulations of the heton's component vortices would cause the heton to rotate in the opposite direction. However, as the curvature is caused only by the AG contribution, the direction of curvature is the same regardless of which component vortex is positive and which is negative,

\subsection{Solid horizontal boundary}
In the oceanographic context, idealized vortex dynamics is typically studied by placing a solid boundary, representing the ocean surface, at $z=0$, where the vertical velocity is zero. This boundary condition can be satisfied for  \qgp point vortices with arbitrary $N$ using the method of images. Each physical vortex below the surface is supplemented by an image vortex at the same horizontal location and equally spaced above the surface, i.e., vortex $i$ at $\bx_i = (x,y,z)$ has an image vortex at $\bx_{i_{image}} = (x,y,-z)$. Then it can be seen from the equations of motion that the pair-vortex vertical velocity induced on the surface by a vortex and its image is zero. Furthermore, the pair-vortex vertical velocity at the surface induced by vortex $i$ and the image of vortex $j$ is nonzero but is cancelled by the vertical velocity induced by the image of vortex $i$ and vortex $j$. To keep the images vertically aligned with their physical vortices as the vortices evolve, their horizontal velocities must be equal, requiring $\Gamma_{i_{image}}=\Gamma_i$. Due to the $1/|\bx|^3$ scaling of the ratio of the AG and QG velocities, the AG contribution from image vortices decays rapidly as vortex depth increases. The asymptotic horizon provides a minimum depth for \qgp point vortices with a solid surface.

While image vortices keep the vertical velocity at the boundary zero, they do, in general, impact the vertical velocity below the boundary. Two physical vortices corresponds to a system with $N=4$, which will, in general, have vertical vortex advection. However, due to the symmetry of the images, the vertical advection of two physical vortices with a solid horizontal boundary is zero. Each vortex, vortex 1, say, has a vertical velocity from three pair-vortex interactions, $2$ and $2_{image}$, $2$ and $1_{image}$, and $1_{image}$ and $2_{image}$. Inspection of the dynamical equations shows that the advecting vertical velocity for each of these pairs is zero. Thus a solid horizontal boundary does not induce vertical vortex motion in a system of two physical vortices.

\section{Interesting \qgp point vortex numerical solutions}

\label{sec:trajectories}

In both 2d and QG, there is a large literature of  point vortex and passive tracer solutions. A detailed study of \qgp trajectories is beyond the scope of this work. Here we choose a small number of configurations and present numerical simulations that highlight some of the new features of \qgp point vortex dynamics. All trajectories presented here are outside the asymptotic horizon. All \qgp simulations are performed with $\Ro = 0.2$. 

\subsection{Passive particle motion induced by a same-sign vortex pair}
\label{sec:passives}

As discussed above, \qgp same-sign vortex pairs have no vertical vortex advection and co-rotate in horizontal planes. 
Passive tracer trajectories in the field of a \qgp same-sign pair can experience vertical advection if the vortices are tilted, while if the vortices are vertically aligned, the vertical advection is zero. Point vortices, being non-diffusive, technically stir the fluid; mixing requires diffusion. Point-vortex stirring does, however, give a good indication of the mixing that would ensue in a vortex-gas with small diffusion. A detailed study of stirring would require investigating the complex three-dimensional nature of passive particle trajectories. Here we focus on the simpler vertical excursion of passive particles, the difference between the maximum and minimum height of a particle along its trajectory. This simpler measure already shows interesting behavior.

\begin{figure}
  \centering
    \includegraphics[width=12cm]{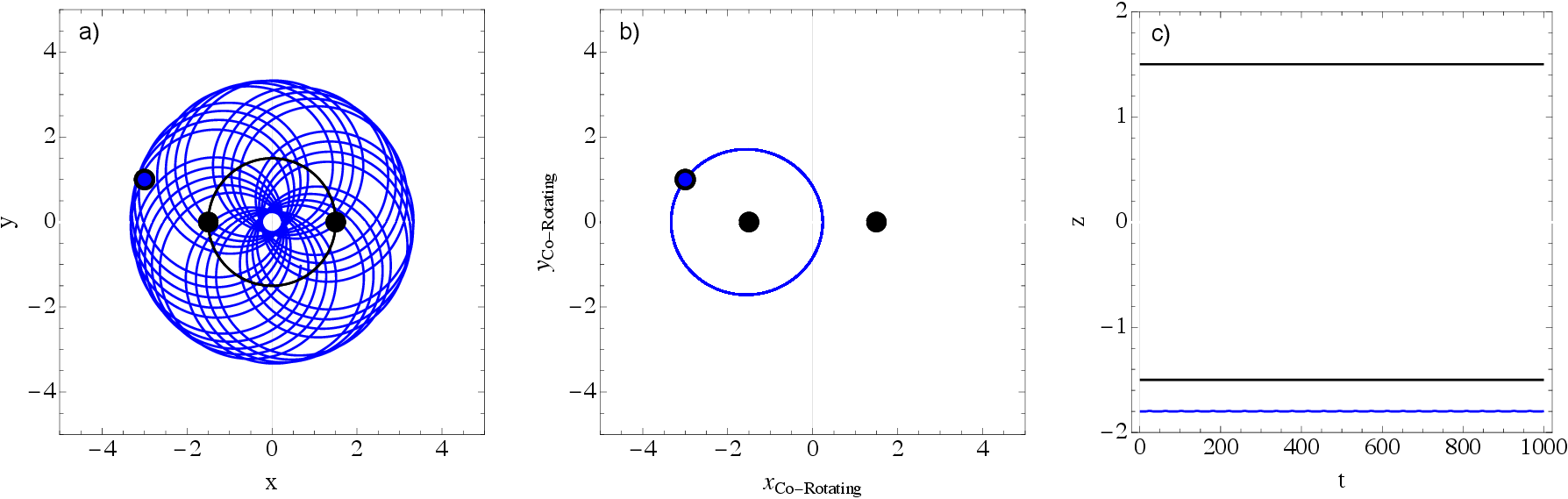}
  \caption{Vortex and passive particle trajectories for a pair of same-sign vortices starting at $\pm(1.5,0,1.5)$ with $\Gamma_1 = \Gamma_2 = 4\pi$, a passive particle starting at $(-3.0, 1.0, -1.8)$, $\Ro = 0.2$, and $0 \le t \le 1000$. a) Horizontal projection of vortex trajectories (black line), passive trajectory (blue line), initial vortex locations (black circles) and initial passive location (blue circle). b) Projection of the trajectories and initial positions onto horizontal coordinates co-rotating with the vortices. c) Time series of the vortex and passive heights, $z(t)$.}
  \label{fig:SSPassiveTraj1}
\end{figure}

\begin{figure}
  \centering
    \includegraphics[width=12cm]{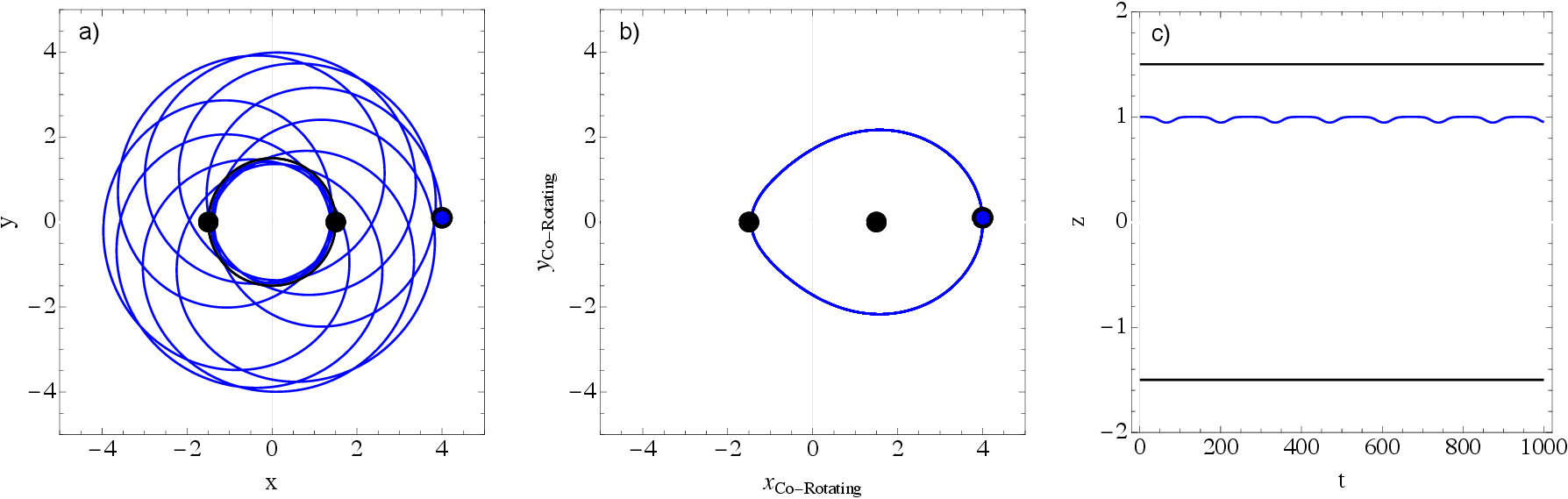}
  \caption{Same as figure~\ref{fig:SSPassiveTraj1} for a  passive particle starting at $(4.0, 0.1, 1.0)$.}
  \label{fig:SSPassiveTraj2}
\end{figure}

\begin{figure}
  \centering
     \includegraphics[width=12cm]{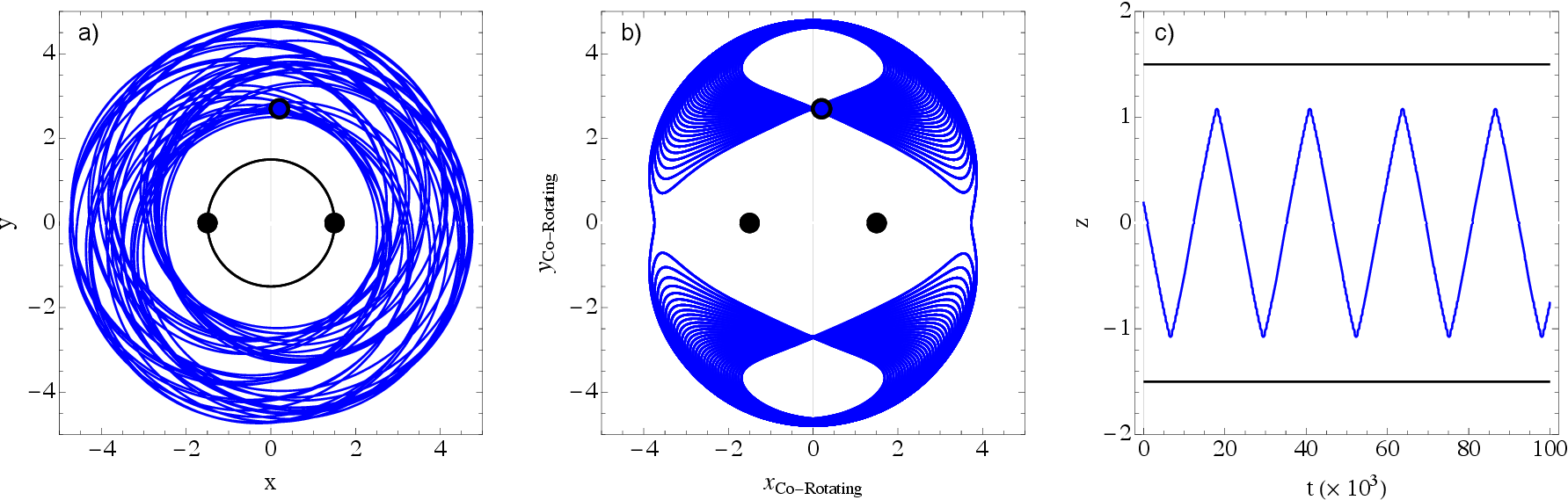}
  \caption{Same as figure~\ref{fig:SSPassiveTraj1} for a passive particle starting at $(0.2, 2.7, 0.2)$ but with the time span in panel a) being $0 \le t \le 10,000$ and the time span in panels b) and c) being $0 \le t \le 100,000$.}
  \label{fig:SSPassiveTraj3}
\end{figure}

We consider a tilted same-sign vortex pair with equal circulations $\Gamma_1 = \Gamma_2 = 4\pi$ at locations $\bx = \pm(1.5, 0, 1.5)$ and with $\Ro = 0.2$. The period $\tau$ of QG co-rotation is $\tau_{QG} \approx 240$. For this configuration, the AG correction slows the vortices down and the \qgp co-rotation period is $\tau_{QG^{+1}}\approx 242$.

Figs.~\ref{fig:SSPassiveTraj1}-\ref{fig:SSPassiveTraj3} show trajectories from three different passive particle initial conditions chosen to display a variety of behaviors. The horizontal motion (panel a) looks quasiperiodic rather than chaotic. In a frame co-rotating with the vortices the motion becomes simpler and appears periodic (panel b). The initial condition of the passive particle in figure~\ref{fig:SSPassiveTraj3} is particularly interesting as it displays large slow vertical excursions. For this third initial condition, the upward (downward) moving branches of the trajectory occur with $y_{Co-Rotating} < 0$ ($y_{Co-Rotating} > 0$), and the height extrema occur as the trajectory crosses the $y_{Co-Rotating}=0$-axis.

\begin{figure}
  \centering
    \includegraphics[width=6cm]{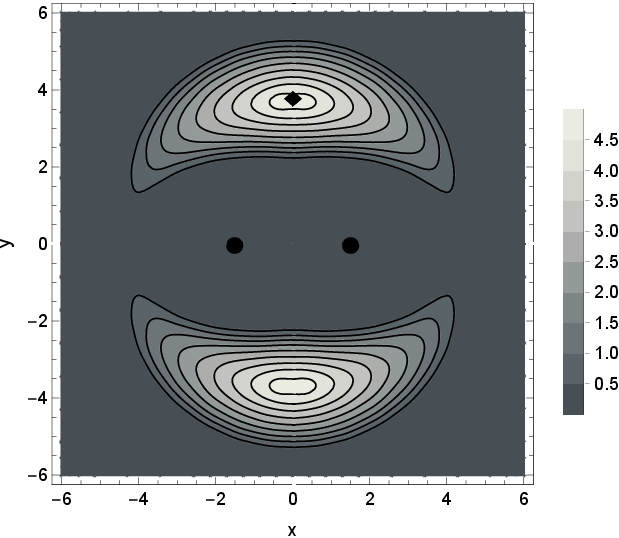}
  \caption{Vertical excursion, $\hbox{Max}(z(t)) - \hbox{Min}(z(t))$ of passive particles as a function of their initial position $(x,y,0)$. The black circles indicate the initial horizontal position of the vortices at $\pm (1.5, 0, 1.5)$, with $\Gamma_1 = \Gamma_2 = 4\pi$ and $\Ro = 0.2$.. The black diamond is the location of the line of passive initial conditions shown in figure~\ref{fig:zline}}
  \label{fig:zexcursion}
\end{figure}

\begin{figure}
  \centering
     \includegraphics[width=6cm]{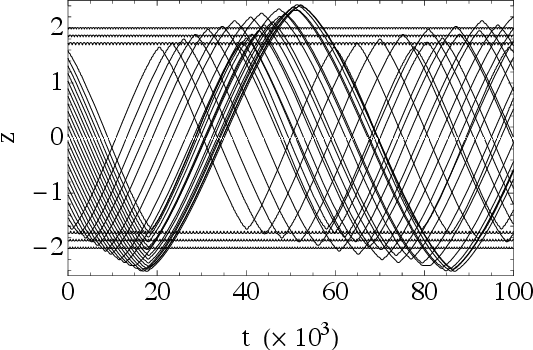}
  \caption{Height vs time of passive particles in the field of a same-sign vortex pair. The vortices are initially at $\pm(1.5,0,1.5)$, have $\Gamma_1 = \Gamma_2=4 \pi$, and $\Ro = 0.2$. The  initial conditions of the passive particles are equally spaced in $z$ along a vertical line with $(x,y) = (0,3.82)$, indicated by the black diamond in figure~\ref{fig:zexcursion}. }
  \label{fig:zline}
\end{figure}

We explore these large vertical excursions by looking at the maximum vertical excursion of passive particles starting on the midplane of the vortices $z=0$ (figure~\ref{fig:zexcursion}). The symmetry of the figure reflects the symmetry of initial vortex positions and the symmetry of \qgp dynamics. One sees that there is an $O(1)$-sized region of midplane passive initial conditions with $O(1)$ vertical excursions larger than the vertical vortex separation. 

We next look at the vertical excursion of passives with initial conditions along a vertical line, $\bx(0) = (0, 3.82, z_0)$ going through q point near the maximum of figure~\ref{fig:zexcursion} and denoted by a black diamond in that figure. The heights (figure~\ref{fig:zline}) show that the region of large vertical excursions extends vertically across the entire vortex separation. This suggests that there are three-dimensional $O(1)$-sized lobes that are stirred in the vertical. 
Trajectories starting on this line appear to fall into two categories: those with $O(1)$ vertical excursions on long  timescales, and those with very small vertical excursions and no long timescales. 

\subsection{Three same-sign vortices}
\label{sec:threess}

Here we investigate the motion of a system composed of three same-sign vortices. We make no attempt to fully explore the large parameter space and instead focus on a few configurations which have interesting behavior. The three vortices have identical circulations, $\Gamma_i = 4\pi$ and are equally distributed in the vertical coordinate. The upper and lower vortices' initial positions lie along a line in the $x-z$ plane, at positions $\pm(1.5, 0, 1.5)$. The middle vortex initially lies on the midplane, offset in the $y$-direction, at $(0, y_0,0)$.

We first consider an initial condition where all three vortices lie along a line, i.e. $y_0 = 0$. For this configuration, both QG and \qgp point vortex dynamics are relatively simple due to the symmetry of the initial condition. The induced velocities of the upper and lower vortices are horizontal and these vortices co-rotate. The middle vortex is stationary. The QG period of rotation for the upper and lower vortices is $\tau_{QG}=48.0$ and for \qgp vortices with $\Ro = 0.2$, $\tau_{QG^{+1}} = 52.3$. 

\begin{figure}
  \centering
     \includegraphics[width=12cm]{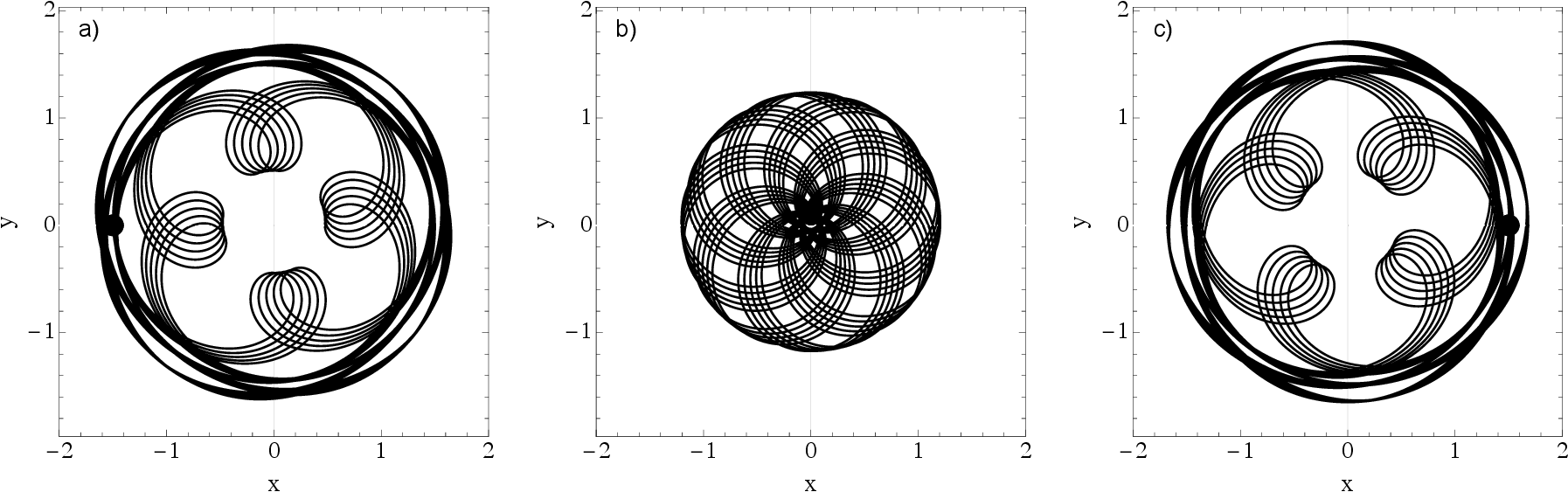}
  \caption{Trajectories of a system of three QG point vortices, $\Ro = 0.0$, initially at $\pm (1.5, 0, 1.5)$ and $(0,y_0,0)$ with $y_0=0.1$. The circulations are  equal, $\Gamma_i = 4\pi$. a) Horizontal projection of the trajectory of the upper vortex for $0 \le t \le 1500$. The black circle indicates the initial position. b) Same as panel a) for the middle vortex. c) Same as panel a) for the lower vortex.}
  \label{fig:threessqg0smally0}
\end{figure}

\begin{figure}
  \centering
    \includegraphics[width=12cm]{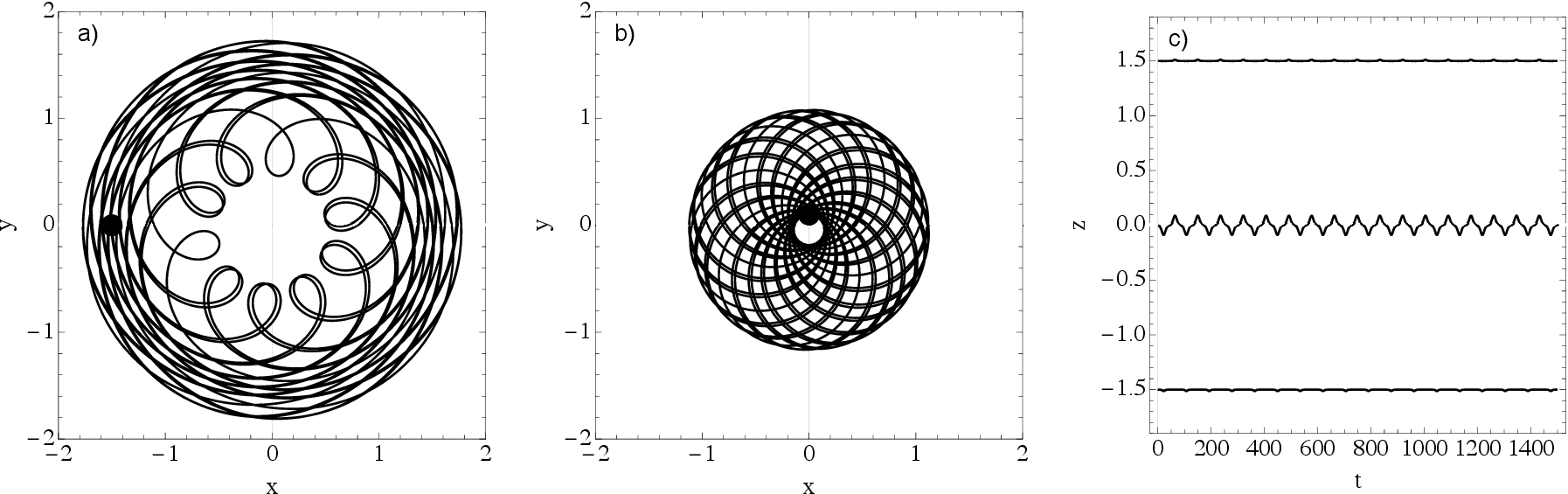}
  \caption{Trajectories of a system of three \qgp point vortices with $\Ro = 0.2$, initially at $\pm (1.5, 0.1, 1.5)$ and $(0,y_0,0)$ with $y_0 = 0.1$. The circulations are equal, $\Gamma_i = 4\pi$. a) Horizontal projection of the trajectory of the upper vortex for $0 \le t \le 1500$. The black circle indicates the initial position. b) Same as panel a) for the middle vortex. The horizontal projection of the lower vortex trajectory is qualitatively similar to panel a). c) Height vs. time of the three vortices.}
  \label{fig:threessqgpsmally0}
\end{figure}

Next consider an initial condition with the middle vortex slightly offset from the origin $y_0=0.1$. The QG trajectories appear quasiperiodic (figure~\ref{fig:threessqg0smally0}). The horizontal projection of the \qgp trajectories also appear quasiperiodic but vortices now also disply small vertical oscillations with periods similar to the horizontal rotation timescale.  (figure~\ref{fig:threessqgpsmally0}).

\begin{figure}
  \centering
     \includegraphics[width=12cm]{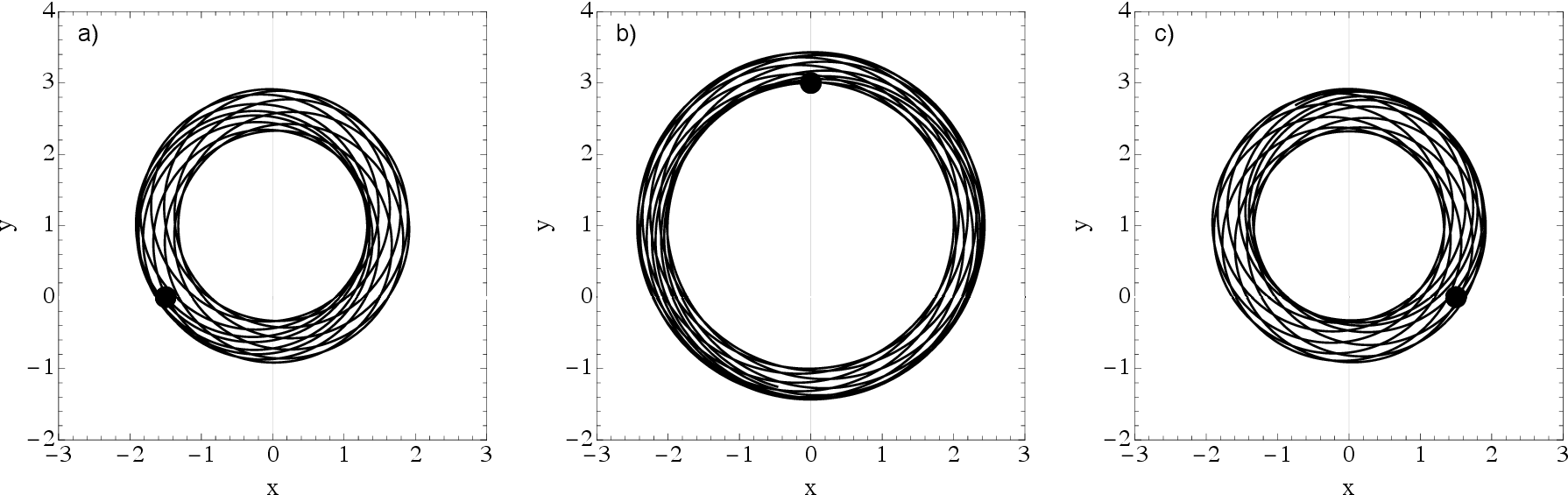}
  \caption{Same as figure~\ref{fig:threessqg0smally0} (three QG point vortices) but with $y_0=3$.}
  \label{fig:threessqg0largey0}
\end{figure}

\begin{figure}
  \centering
     \includegraphics[width=12cm]{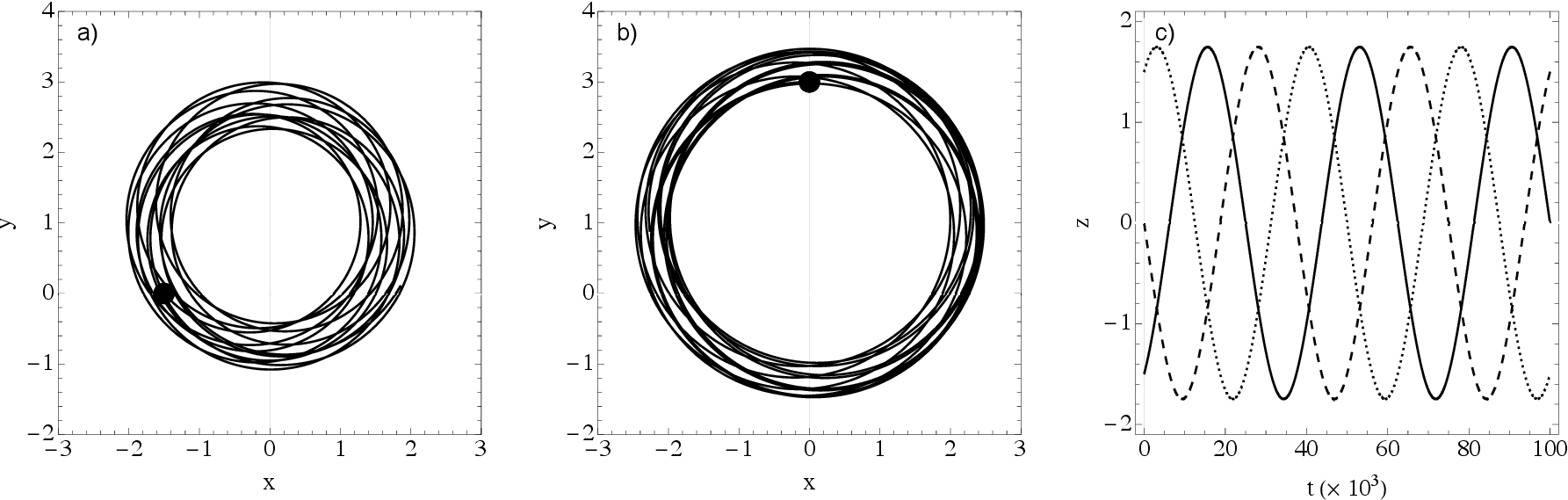}
  \caption{Same as figure~\ref{fig:threessqgpsmally0} (three \qgp point vortices) but with $y_0 = 3$.}
  \label{fig:threessqgplargey0}
\end{figure}

Now consider a system where the middle vortex initially has a significantly larger horizontal offset, $y_0 = 3$ (figure~\ref{fig:threessqg0largey0}). The QG dynamics still looks quasiperiodic. The \qgp trajectories for this initial condition display similar horizontal trajectories as in the QG case for the time shown, a few horizontal rotation times. The notable new feature is the large $O(1)$ vertical oscillation which causes the vortices to exchange their vertical positions. The vertical motion appears periodic with a period $O(100)$ times longer than the horizontal rotation timescale. As the vortex heights slowly change, the shorter timescale horizontal motions change their character slowly but remain qualitatively similar.

\begin{figure}
  \centering
     \includegraphics[width=8cm]{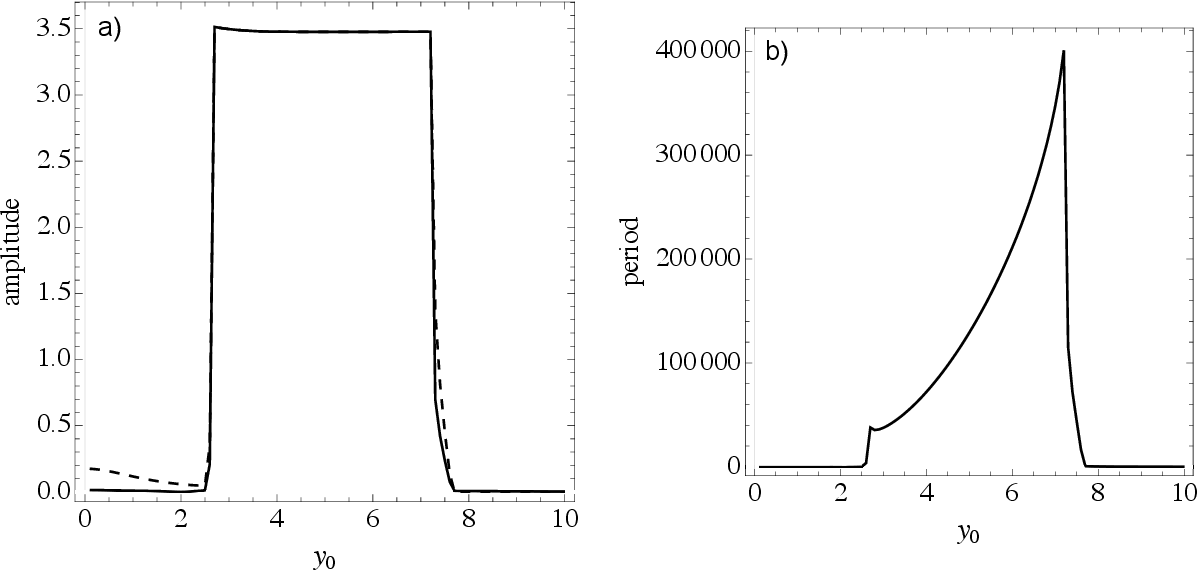}
  \caption{Amplitude (panel a) and period (panel b) of the vertical oscillation of the vortex initially at $z=0$ in a system of three \qgp point vortices initially at $\pm (1.5, 0, 1.5)$ and $(0, y_0, 0)$ as a function of the offset $y_0$. The vortices have $\Gamma_i = 4\pi$ and $\Ro = 0.2$. In panel a), the solid line is the amplitude of the initially upper and lower vortices and the dashed line is the amplitude of the initially middle vortex. The periods of the three vortices are identical.}
  \label{fig:threessqgpampsperiods}
\end{figure}

The vertical oscillations have an interesting structure as a function of the initial condition offset $y_0$ (figure~\ref{fig:threessqgpampsperiods}). There is a region of $y_0$ with large vertical oscillations similar to those seen above for $y_0=3$. Note that this region is similar in location to the region found in the previous section where passive particles in the field of two vortices have large vertical excursions, but here the boundary of the region appears sharper. The amplitude of the vertical oscillation of the three vortices are qualitatively similar but with small quantitative differences. The periods of the oscillation of the three vortices are identical at each $y_0$.

\subsection{chaotic motion}
\label{sec:chaos}

\begin{figure}
  \centering
     \includegraphics[width=12cm]{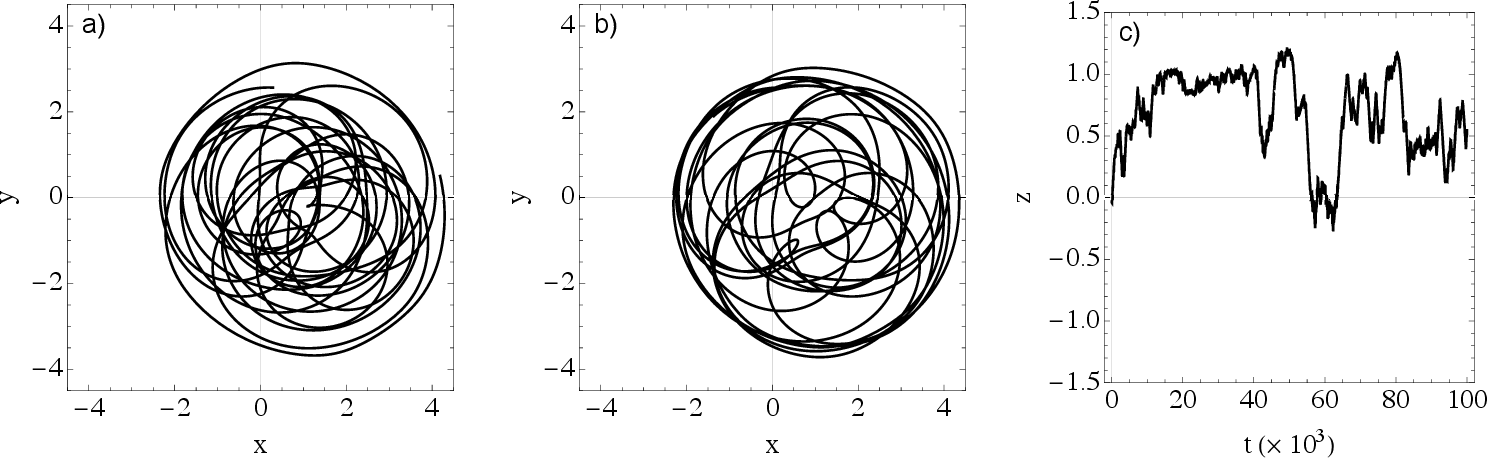}
  \caption{Trajectories of a system of 4 vortices with $\Gamma_i = 4\pi$ with positions chosen at random. a) Horizontal motion of one vortex under QG dynamics, $\Ro = 0$, for $0 \le t \le 1500$. b) Horizontal motion of the same vortex under \qgp dynamics starting from the same initial condition with $\Ro = 0.2$. c) Vertical motion of this \qgp vortex.}
  \label{fig:chaos}
\end{figure}

The above trajectories look quasiperiodic. Their Lyapunov exponents appear to converge to zero as the integration time grows. They are most likely not chaotic. However, like 2d and QG point vortices, \qgp point vortices do have chaotic motion. Four-vortex configurations can be found with chaotic trajectories with nonzero Lyapunov exponents (figure~\ref{fig:chaos}). The horizontal trajectories of the chaotic \qgp vortices are qualitatively similar to that of their corresponding chaotic QG vortices. For this chaotic trajectory, the \qgp vortices have $O(1)$ irregular vertical motion on long timescales.

\section{Discussion}

We have developed a new class of point vortex dynamics, three dimensional ageostrophic balanced point vortices. The solutions are found as point vortex idealizations of vortex gas solutions of a specific set of asymptotic balance equations, the \qgp equations. The inclusion of ageostrophic dynamics in rapidly rotating, stably stratified flows brings both qualitative and quantitative differences. Horizontal velocities are perturbed by small, order Rossby number, corrections to QG horizontal flow. Vertical velocities are also small but as they are perturbations from zero, they produce qualitatively new phenomena. 

We have seen that some configurations of \qgp point vortices, and of passive particles in the field of \qgp point vortices, have large $O(1)$ vertical transport. As the vertical velocity is small, large vertical transport requires correspondingly long times. We have seen that the regions of initial conditions with large vertical transport can be $O(1)$ in size and these regions can have relatively sharp boundaries. Vertical transport and mixing are important processes in rotating stratified turbulent flow and we expect that \qgp point vortex dynamics will prove useful in understanding ageostrophic vertical transport and mixing across many geophysical and astrophysical settings.

It is well known that 2d and QG point vortex dynamics have  Hamiltonian structures which provide insight into the dynamical behavior of the vortices. This fact, together with the behavior of the trajectories seen here, leads us to think it likely that \qgp point vortex dynamics is Hamiltonian. There is a large literature on Hamiltonian fluid dynamics, e.g., \cite{Morrison1998}. Based on the theoretical understanding of Hamiltonian fluid dynamics, we expect that the imposition of balance and the asymptotic expansion in \qgp will preserve the Hamiltonian structure present in the primitive equations. We also note that the balance formulation of \citep{Holm1996} has an explicit Hamiltonian structure. We leave the development of any potential Hamiltonian structure in \qgp point vortices for future work. 

\backsection[Acknowledgements]{The author would like to acknowledge fascinating and useful discussions with Jean-Luc Thiffeault, Philip J. Morrison, and Charles R. Doering.}

\backsection[Funding]{This work was performed in part at Aspen Center for Physics, which is supported by National Science Foundation grant PHY-1607611. This work utilized resources from the University of Colorado Boulder Research Computing Group, which is supported by the National Science Foundation (awards ACI-1532235 and ACI-1532236), the University of Colorado Boulder, and Colorado State University.}

\backsection[Declaration of interests] {The authors report no conflict of interest.}

\backsection[Data availability statement]{Mathematica code to calculate the far-field equations in section~\ref{sec:pvdynamics} is available at https://github.com/JeffreyWeiss/QGplus1PointVortices.}

\backsection[Author ORCID]{J.B. Weiss,
https://orcid.org/0000-0002-0706-861X}

\bibliographystyle{jfm}

\bibliography{references}

\end{document}